# DIAMETRICAL MESH OF TREE (D2D-MoT) ARCHITECTURE: A NOVEL ROUTING SOLUTION FOR NoC

Prasun Ghosal*, Sankar Karmakar

**Address for Correspondence**
Department of Information Technology, Bengal Engineering and Science University, Shibpur
Howrah 711103, WB, India

**ABSTRACT**
Network-on-chip (NoC) is a new aspect for designing of future System-On-Chips (SoC) where a vast number of IP cores are connected through interconnection network. The communication between the nodes occurred by routing packets rather than wires. It supports high degree of scalability, reusability and parallelism in communication. In this paper, we present a Mesh routing architecture, which is called Diametrical 2D Mesh of Tree, based on Mesh-of-Tree (MoT) routing and Diametrical 2D Mesh. It has the advantage of having small diameter as well as large bisection width and small node degree clubbed with being the fastest network in terms of speed. The routing algorithm ensures that the packets will always reach from source to sink through shortest path and is deadlock free.

## I. INTRODUCTION
*A. Introduction to Network-on-chip*
Present day system-on-chip (SoC) design contains billions of transistors. One of the major problems associated with future SoC design comes up from non-scalable global wire delays. Global wires carry signals across a chip but these wires do not scale in length with technology scaling. In ultra-deep-sub-micron processes, 80% or more of the delay of critical paths will be due to interconnects. Secondly, for a long bus line, the intrinsic parasitic resistance and capacitance can be quite high. If the bus length increases and/or the number of IP core blocks are increased then the associated delay in bit transfer over the bus become arbitrarily large and exceeds the targeted clock period. Thirdly, the power consumption increases with the circuit size. Finally, In SoC, a bus allows only one communication at a time, so all buses of the hierarchy are blocked, as its bandwidth is shared by all the system attached to it.

To overcome all the above mentioned limitations we consider the architecture of Network-on-Chip (NoC) [3]. NoC is a new electronic device for designing future SoCs where various IP cores are connected to the router based network. The network is used for packet switched on-chip communication among cores [2] [3] [4] [5] [6] [7].

*B. Basic Components of NoC Routing*
The basic components of NoC routing consist of three fundamental building blocks viz. (i) Switch, which are called as routers, (ii) The Network Interfaces (NI) which are also called network adapters, and (iii) The last one is Link. The components are shown in Figure 1.

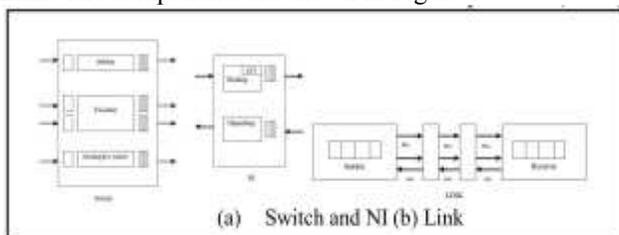

**Figure 1: Basic components in NoC routing**

The backbone of the NoC consists of switches, whose main function is to route the packets from source to destination. Some NoC design depends on octagon or ring connectivity. This provides the logical control. NoC can be based on circuit or packet switching, or combination of both.

An NI connects each core to the NoC. NIs convert transactions of requests/responses into packets and vice versa. Packets are split into a FLow control unITS called as *FLITS* before transmission. The Look Up Table (LUT) specifies the path that packet will follow inside the network and reach the destination [7].

## II. PROBLEMS IN NOC ROUTING
Design of an NoC consists of several problem areas [3]. These are as follows.
1. The topology synthesis problem.
2. The channel width problem.
3. The buffer sizing problem.
4. The floor-planning problem.
5. The routing problem.
6. The switching problem.
7. The scheduling problem.
8. The IP mapping problem.

Among these the most important problem in NoC design is routing problem. The network performance and power consumption are greatly affected due to this phase only.
A basic routing problem in NoC may be stated like this:
***Input:*** *An application Graph, a communication architecture A(R,ch), the source and destination routers.*
***Find:*** *A decision function at router r, RD (r,s,d,ρ(n) for selecting an output port to route the current packet(s) while achieving a certain objective function.*

There may be different approaches to solve the current problem. Two things, one is the complexity of implementation and another is performance requirement, are the most considerable during solving the routing architecture design. Compared to the adaptive routing, deterministic routing is mostly useful over the uses of less resources and guarantee to arrival packets. But adaptive routing algorithms give better throughput.

## III. EXISTING ARCHITECTURES IN NOC
Different topologies have been developed for NoC architecture [3] [7]. Some are as follows.
*A. Cliché topology*
This architecture consists of an M × N mesh of switch, the edges is connected four neighboring switches and one IP block. It has M rows and N columns, Diameter: (M+N-2), Bisection Width: min(M,N), Number of Routers required: (M × N), Node Degree: 3(corner), 4(boundary), 5(central). This is shown in Figure 2.

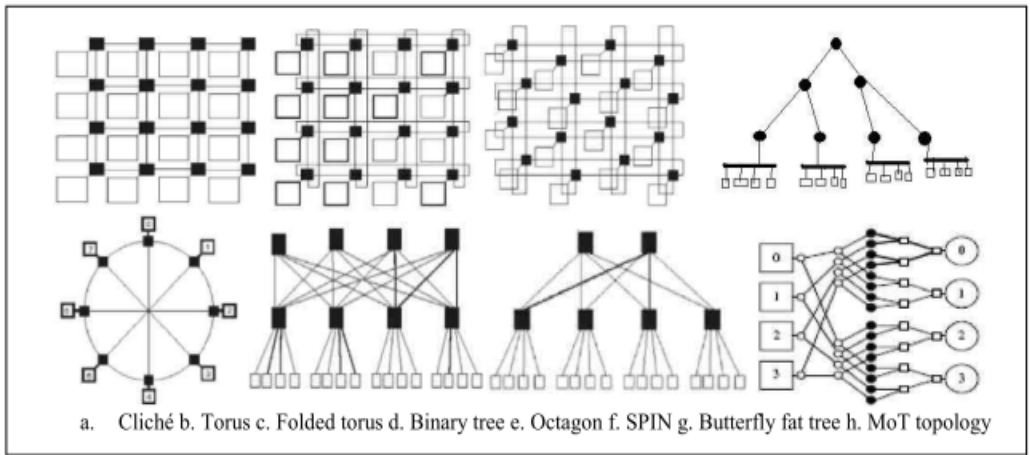

**Figure 2: Different topologies**

*B. Torus Topology*

In torus architecture [14], the only difference with mesh is that the switches at the edges are connected to the switches at the opposite edge through wrap-around channels. It has also M rows and N columns, Diameter: [M/2] + [N/2], Bisection Width: 2 × min (M,N), No of Routers required: (M × N), Node Degree: 5(five). This is shown in Figure 2.

*C. Folded Torus*

In folded Torus the only difference from Torus is that the long end-around connections are avoided by folding the Torus, to avoid the excessive delay due to long wires, for folded Torus having M rows and N columns. Diameter: [M/2] + [N/2], Bisection Width: 2 × min (M,N), No of Routers required: (M × N), Node Degree: 5(five). This is shown in Figure 2.

*D. Binary Tree*

In Binary tree (Figure 2) 4 IP cores are connected at the leaf level node, but none at the others. A Binary tree based network with N number of IP core has Diameter: $\log_2 N$, Bisection Width: 1, No of Routers required: (N-1), Node degree: 5(leaf), 3(stem), 2(root).

*E. Octagon*

Each node in this network (Figure 2) is associated with an IP and a switch. For a system consisting of more than eight nodes, the network is extended to multidimensional space. For a network having N number of IP blocks, Diameter: 2[N/8], Bisection width: 6 for N≤8 or 6(1+[N/8]) for N>8, No of router required: 8 for n≤ 8 or 8 and (1+[N/8])−[N/8] for N>8. Node Degree: 4(member node), 7(bridge node).

*F. SPIN*

Every node has four children and the parent is replicated four times at any level of the tree, the functional IP blocks reside at the leaves and the switches reside at the vertices, for N number of IP blocks the network has, Diameter: $\log_2 N$, Bisection Width: N/2, Number of router needed: $N\log_2 (N/8)$, Node Degree: 8(non root), 4(root). This is shown in Figure 2.

*G. Butterfly Fat Tree*

In this network (Figure 2), IPs are placed at the leaves and switch placed at the vertices. For N number of IPs the network has Diameter: $\log_2 N$, Bisection Width: √N, Number of router needed N/2, Node Degree: 6(non root), 4(root).

*H. MoT (Mesh of Tree)*

A 4 × 4 MoT network (shown in Figure 2) consist of 4 row trees and 4 column trees. The PCs act as sources for packets and MMs are at the roots of the trees. Each terminal of MoT network could serve as a processor cluster up to 16 processors. For N × N MoT the network has Diameter: $4 \log_2 N$, Bisection width: N, Number of routers required $(3N^2 - 2N)$, Node Degree : 2(leaf), 3(stem), 18(root).

An M × N MoT [11] [12] where M and N denote the number of row and column trees has

1. Number of nodes = 3 × (M × N) – (M+N)
2. Diameter = 2 $\log_2 M$ + 2 $\log_2 N$
3. Bisection width = min(M,N)
4. Recursive structure
5. A maximum of two layers, horizontal and vertical, are sufficient for routing.

*1) Addressing*

The address of routers in M × N MoT consists of four fields.

1. Row Number
2. Column Level
3. Column Number
4. Row Level

The details of the addressing scheme may be omitted due to paucity of space.

*2) Routing Algorithm*

The routing algorithm given here is the deterministic routing approach. The routing algorithm ensures that the packet will reach to destination always through specified shortest path. We use the following abbreviations to describe the algorithm. Let RN: Row Number, CL: Column Level, CN: Column Number, RL: Row Level, addr (curr): address of the current node, addr (dest): address of the destination node. Each router executes the same algorithm as proposed in Figure 3.

```
Route-MoT()
If (RN of addr (curr) ≠ RN of addr (dest))  // Step1
Route to Column Parent;
Else If (CL of addr (curr) ≠ CL of addr (dest))  // Step2
Route to Column Child having equal RN as addr (dest);
Else If (CN of addr (curr) ≠ CN of addr (dest))  // Step3
Route to Row Parent;
Else If (RL of addr (curr) ≠ RL of addr (dest))  // Step4
Route to Row Child having equal CN as addr (dest);
Else If (Destination Core-ID field = 0)  // Step5
Route to Core1;
Else Route to Core2;
```

**Figure 3: MoT routing algorithm**

*I. Diametrical 2D Mesh*

Diametrical 2D Mesh is a performance efficient topology [13], because its network diameter is reduced considerably in comparison to 2D Mesh. Although, in this network the area is increased because of 8 extra links, its power consumption is decreased due to average hop count reduction. Number of extra links in Diametrical 2D Mesh do not grow by the growth of IP cores. On the other hand, by growth of IP cores, number of extra links constantly and statistically equals 8. This link redundancy is decreased by the growth of IP cores. For example, the link redundancy in 16 IP cores 2D Mesh is 1/3 and in 25 IP cores 2D Mesh, this ratio is 1/5. Furthermore, the network diameter is decreased 50% with any number of IP cores in comparison to 2D Mesh.

All in all, the main purpose of adding 8 extra links to 2D Mesh topology is the reduction of diameter in 2D Mesh when 2D Mesh is expanded by large number of IP cores. Moreover, it was tried to minimize the area and power consumption redundancy by defining constant 8 links that decrease diameter and connect four edge sub-networks.

*1) Diametrical 2D Mesh Addressing*

Diametrical 2D Mesh uses addresses those are composed of two parts; X and Y, which X shows the number of row and Y indicates the number of column that the node is located in (see Figure 4). The number of bits in X and Y parts are determined by the number of rows and columns in Diametrical 2D Mesh. For example, for 16 IP cores, Diametrical 2D Mesh has 4 nodes in each rows and 4 ones in each column, due to this fact both X and Y parts have to use 2 bits. In other words, 2 bits show the number of row, X, and 2 bits indicate the number of column, Y. Also, for 25 IP cores, Diametrical 2D Mesh has 5 nodes in each row and column. Therefore, both X part bits and Y part bits must be three. This means that 6 bits addresses are used to label the Diametrical 2D Mesh with 25 IP cores. Both 16 and 25 Diametrical 2D Mesh addressing are demonstrated in Figure 4.

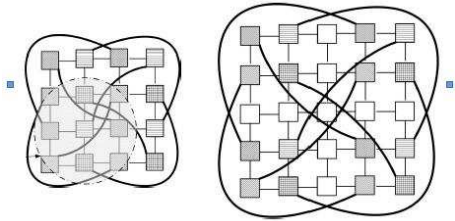

**Figure 4: Diametrical 2D mesh architecture**

*2) Routing in Diametrical 2D Mesh*

The routing protocol deals with resolution of the routing decision made at every router. Routing method affects the cost (area and power consumption) and the performance (average latency and throughput) issues in the NoC design.

We propose extended XY routing for diametrical 2D mesh. Fundamentally, this routing algorithm is the shortest path and inherited from the well-known 2D Mesh XY routing. Similar to XY routing, Extended XY is very simple due to simple addressing scheme and structural topology. Figure 3 shows the details of the routing pseudo code. As can be seen, we define $X_{offset}$ and $Y_{offset}$ values in the pseudo code, which are calculated as follows.

$$X_{offset} = X_{dest} - X_{current}$$
$$Y_{offset} = Y_{dest} - Y_{current}$$

Where $X_{current}$ is the X value of a current node and $X_{dest}$ is the X value of a destination node. In addition, $Y_{current}$ is the Y value of a current node and $Y_{dest}$ is the Y value of a destination node. $X_{offset}$ and $Y_{offset}$ are the values indicating the number of rows and columns between a current and a destination node respectively. If both $X_{offset}$ and $Y_{offset}$ values are zero, it means that the current node is the destination and the packet reaches the destination node. Extended XY routing utilizes conventional XY routing in the following conditions.

Case I: When the diameter channel is not used:
- If a current node and a destination node are located in the same row or column, or $X_{offset} + Y_{offset} < d-1$, according to these conditions, conventional XY routing decisions are performed.

Case II: When diameter channel is used:
If $X_{offset} + Y_{offset} > d-1$:
- If a source switch has a diameter link, the flits are forwarded via a diameter channel.
- If a source switch has not a diameter channel then firstly, based on XY routing the flits are forwarded to the nearest intermediate node which has a diameter, and, secondly, the flits are forwarded via a diameter channel.

**IV. PROPOSED DIAMETRICAL 2D MESH OF TREE (D2D-MOT) ROUTING ARCHITECTURE**

With the help of Diametrical 2D-Mesh topology and MoT(Mesh of Tree) topology proposed earlier, the new architecture is formed which we called Diametrical 2D Mesh of Tree (D2D-MoT) routing architecture for NoC. Here 4 × 4 rows and column trees are used to form the architecture. The leaf level nodes are common to both the trees. In figure L, S and R denote the Leaf, stem and root level nodes. All these nodes are replaced by routers in practical. At root level, the router is attached to IP cores. Here each leaf node is connected diagonally in same module. There is a stem node between two leaf nodes which is connected. Also, between two stem nodes, there is a root node, which is connected. There are eight root nodes, four roots are external and four are internal, these four internal root nodes are connected oppositely. It has ten extra links that increases the wire length, but reducing the extra hops, and thereby increasing the network performance due to high speed of operation. Also, the Diametrical 2D bypasses channels, that causes around 50 percent reduction in network diameter. Hence, the design of the proposed architecture provides a balanced improvement over the performance as well as cost of the routing in an NoC.

The proposed architecture is shown in Figure 5.

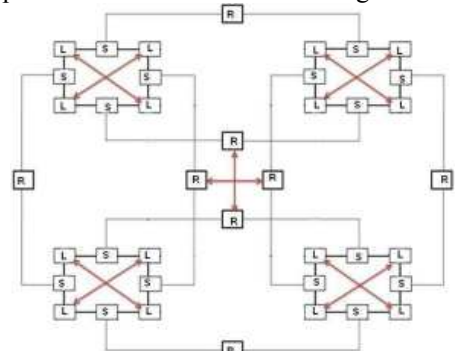

**Figure 5: Proposed diametrical 2D mesh of tree architecture**

The characteristic parameter values of this network architecture (shown in Figure 5) are presented in Table I.

**TABLE I: PERFORMANCE PARAMETER VALUES OF PROPOSED D2D-MOT ARCHITECTURE**

| Node degree | Leaf node = 5 |
| --- | --- |
|  | Stem node = 3 |
|  | Internal root node = 3 |
|  | External root node = 2 |
| No. of IP blocks | 2*(m*n), here the no. of IP core is 32 |
| No. of nodes | 3*(m*n)(m+n), here the total no. of nodes is 40 |
| No. of links | 3*(m*n)+(8+2), here the total no. of links is 58 |

### V. PROPOSED ROUTING ALGORITHM

The routing algorithm follows the deterministic routing approach. The routing algorithm ensures that the packet will reach to destination always through specified shortest path. Thus the proposed network is always Live lock free. We use the following abbreviations to describe the algorithm. Let RN: Row Number, CL: Column Level, CN: Column Number, RL: Row Level, addr (curr): address of the current node, addr (dest): address of the destination node. Each router executes the same algorithm as proposed in Figure 6.

```
Route D2D-MoT()
1. If (source node and sink node is same row)
2. Route to row parent and row child having equal CN;
3. Else If (source and sink node is same column)
4. Route to row parent and row child having equal RN;
5. Else If (source and sink node is not in same row or same column)
6. Route the diametrical channel;
7. {If (The current node and destination node is not in same row)
8. Route to row parent and interchange the router and route to row child having equal CN;
9. Else (The current node and destination node is not in same column)
10. Route to column parent and interchange the router and route to column child having equal RN;
11. If (current and sink is not in same row or same column)
12. Route the diameter channel};
13. Else If (Destination core ID field = 0)
14. Route to core 1;
15. Else Route to core 2.
```

**Figure 6: Proposed D2D-MoT routing algorithm**

### VI. EXPERIMENTAL RESULT

The proposed algorithm is implemented in a standard desktop environment running Linux operating system on a chipset with Intel Pentium processor running at 3 GHz using GNU GCC compiler. The important sections of two sample runs with different parameters are shown below.

- **Sample run 1**

```
*******************************************
 Adjacency Matrix
*******************************************
Input the no of nodes in the each side is = 3
Total Number of node in n x n Matrix is = 9
No IP Blocks is 18
*******************************************
 Input adjacency matrix is:
*******************************************
 1 1 0 0 0 1 1 1 1
 0 0 1 1 1 0 1 0 1
 0 1 1 1 0 0 0 1 1
 1 1 0 0 0 1 1 1 0
 1 1 1 1 0 0 0 0 0
 0 1 0 1 1 0 0 0 1
 1 1 0 1 1 1 0 0 0
 0 1 1 0 1 1 0 1 1
 1 1 1 0 0 1 0 1 1
```

```
No of 1 in the Matrix = 46
No of link  is = 23
********** XY CO-Ordinate is:*******************
 00 01 02 03 04 05 06 07 08
 10 11 12 13 14 15 16 17 18
 20 21 22 23 24 25 26 27 28
 30 31 32 33 34 35 36 37 38
 40 41 42 43 44 45 46 47 48
 50 51 52 53 54 55 56 57 58
 60 61 62 63 64 65 66 67 68
 70 71 72 73 74 75 76 77 78
 80 81 82 83 84 85 86 87 88
Input starting vertex = 5
Input destination = 7
Shortest path = 5 => 8 => 7
Minimum distance = 2
```

- **Sample run 2**

```
*******************************************
 Adjacency Matrix
*******************************************
Input the no of nodes in the each side is = 4
Total Number of node in n x n Matrix is = 16
No IP Blocks is 32
*******************************************
 Input adjacency matrix is:
*******************************************
 0 1 0 0 1 1 0 0 0 0 0 0 0 0 0 0
 1 0 1 0 1 1 0 0 0 0 0 0 0 0 0 0
 0 1 0 1 0 0 1 1 0 0 0 0 0 0 0 0
 0 0 1 0 0 0 1 1 0 0 0 0 0 0 0 0
 1 1 0 0 1 0 0 1 0 0 0 0 0 0 0 0
 1 1 0 1 0 1 0 0 1 0 0 0 0 0 0 0
 0 0 1 1 0 1 0 1 1 0 0 0 0 0 0 0
 0 0 1 1 0 1 0 1 1 0 0 0 0 0 0 0
 0 0 0 0 1 0 0 0 1 0 0 0 1 0 0 0
 0 0 0 0 1 0 0 1 0 0 1 0 1 1 0 0
 0 0 0 0 1 1 0 0 1 0 1 0 0 1 1 0
 0 0 0 0 0 0 1 0 0 1 0 0 0 1 1 0
 0 0 0 0 0 0 0 1 1 0 0 1 0 1 0 0
 0 0 0 0 0 0 0 1 1 0 0 1 0 1 0 0
 0 0 0 0 0 0 0 0 0 0 1 1 1 0 1 0
 0 0 0 0 0 0 0 0 1 0 1 0 0 1 0 0
No of 1 in the Matrix = 68
No of link  is = 34
***************** XY CO-Ordinate is:***********
 00    01   02   03   04   05   06  07  08  09   010
 011   012  013  014  015
 10    11   12   13   14   15   16  17  18  19   110
 111   112  113  114  115
 20    21   22   23   24   25   26  27  28  29   210
 211   212  213  214  215
 30    31   32   33   34   35   36  37  38  39   310
 311   312  313  314  315
 40    41   42   43   44   45   46  47  48  49   410
 411   412  413  414  415
 50    51   52   53   54   55   56  57  58  59   510
 511   512  513  514  515
 60    61   62   63   64   65   66  67  68  69   610
 611   612  613  614  615
 70    71   72   73   74   75   76  77  78  79   710
 711   712  713  714  715
 80    81   82   83   84   85   86  87  88  89   810
 811   812  813  814  815
 90    91   92   93   94   95   96  97  98  99   910
 911   912  913  914  915
 100  101  102 103 104 105 106 107 108 109 1010 1011
 1012 1013  1014  1015
 110  111   112 113 114 115 116 117 118 119  1110
 1111 1112 1113  1114  1115
 120  121   122 123 124 125 126 127 128 129  1210
 1211 1212 1213  1214  1215
 130  131   132 133 134 135 136 137 138 139  1310
 1311 1312 1313  1314  1315
 140  141   142 143 144 145 146 147 148 149  1410
 1411 1412 1413  1414  1415
```

```
150 151  152 153 154 155 156 157 158 159  1510
1511 1512 1513  1514  1515
Input starting vertex = 1
Input destination = 15
Shortest path = 1 => 6 => 11 => 15
Minimum distance = 3
```

Results of comparison with other existing algorithms [15] are summarized in the following Table II.

**TABLE III: COMPARATIVE EXPERIMENTAL RESULTS**

| # IP blocks | Total Transfer Time (ms) (D2DMoT) | Total Transfer Time (ms) (mesh-tree) | Total Transfer Time (ms) (mesh) | % of Speeding up |
|---|---|---|---|---|
| 18 | 12470 | 15575 | 18195 | 20 |
| 32 | 20374 | 30635 | 35380 | 33 |
| 50 | 28247 | 49235 | 66495 | 42 |
| 72 | 32525 | 67342 | 80527 | 51 |

## VII. CONCLUSION AND FUTURE WORK

In D2D-MoT topology, ten (10) extra links are connected [(i-1) × 2 in D2D-MoT to each diagonally router and four internal router], Due to this the wire length increases a little bit but simultaneously it reduced the diameter to 50% compared to other topology. For this it leads less average hop count and average latency and power consumption. The performance depends on network average latency and throughput. Hence we get good overall performance improvement with this architecture.

The D2D-MoT is really advantageous for large number of IP cores. For small number of network or IP cores there is no such significant improvement of its performance compared to other topologies. In future, we shall try to improve our architecture and algorithm to find out better performance, to improve energy and power consumption of the entire network for large number of IPs. Also, another aim is to perform a more careful analysis and incorporation of the parasitic and leakage effects in the design of ultra-low power NoCs.